\documentstyle[preprint,eqsecnum,aps]{revtex}

\begin{document}
\draft
\preprint{ver1.0}
\title{High frequency ESR investigation on dynamical charge
disproportionation and spin gap excitation in NaV$_2$O$_5$}
\author{H. Nojiri, S. Luther and M. Motokawa}
\address{Institute for Materials Research, Tohoku University, Katahira 2-1-1,
Sendai 980-8577, Japan }
\author{M. Isobe and Y. Ueda}
\address{Institute of Solid State Physics, University of Tokyo, Roppongi Minato-ku,
Tokyo 106-8666, Japan }
\date{\today}
\maketitle
\begin{abstract}
 	A significant frequency  dependence of the ESR line width is found in NaV$_2$O$_5$ between 34-100 K and the line width
increases as the resonance frequency is increased from 95 GHz to 760 GHz.
The observed frequency dependence is qualitatively explained in terms of the dynamical charge disproportionation.
The present results show the essential role of the internal charge degree of freedom in a V-O-V bond. 
We have also proposed the existence of the  Dzyaloshinsky-Moriya interaction in the low temperature charge
ordered phase considering the breaking of the selection rule of ESR realized as the direct observation of the spin gap excitation.

\end{abstract}
\pacs{75.30.Et, 75.40.Gb, 75.30.Kz,75.10}

\narrowtext
\section{Introduction}
\label{sec:level1}

A charge degree of freedom in NaV$_2$O$_5$ and its role for the spin gap formation is the problem of current interest. 
At first, the opening of a spin gap below {\em T}$_c$=34 K was interpreted as the indication of spin-Peierls
transition.\cite{Isobe}
The one-dimensional character observed in the magnetic susceptibility has been
explained by the model, in which magnetic chains made up of V$^{4+}$ ions are running along
the {\em b}-axis and these chains are separated by the non-magnetic chains consists of V$^{5+}$ ions.
The basis of this model is the structural analysis by Carpy and Galy suggesting the noncentrosymmetric space group {\em P}2$_1$mn.\cite{Carpy}
However, it has been reported recently by the X-ray diffraction measurements that only single V site exists at room temperature and the crystal structure
belongs to the centrosymmetric space group {\em P}mmn.\cite{Smolinski,Meetsma,Schnering}
This new structural model shows that NaV$_2$O$_5$ is a mixed valence(MV) compound at room temperature and a spin is delocalized in a V-O-V bond, which is
a rung of a quarter-filled two-leg ladder as shown in Fig.1.
Hence, the origin of the transition at {\em T}$_c$ has been reconsidered as the charge ordering(CO) and several
theoretical models have been proposed.\cite{Seo,Thalmeier,Nishimoto,Mostovoy,Gros}
In fact, the evidences of CO have been reported in NMR\cite{Ohama}, thermal conductivity and dielectric constant.\cite{VasilevT,Smirnov,Sekine}

An important nature of the CO of NaV$_2$O$_5$ is that many kinds of thermodynamic quantities develop continuously below {\em T}$_c$ and that a
first-order nature is almost absent.\cite{Ravy}
As is well known, for a second-order phase transition, we can expect a short-range correlation of the order parameter above {\em T}$_c$.
The second-order like behavior in NaV$_2$O$_5$ implies that a short range correlation of the charge develops toward the static CO at {\em T}$_c$.
In fact, precursor phenomena of the transitions are observed in many experiments such as, optical conductivity\cite{Damascelli}, sound velocity\cite{Fertey} and
X-ray diffuse scattering\cite{Ravy}.
At this point, it is worth reminding us another important point that the CO in NaV$_2$O$_5$ is unconventional in a sense that the order parameter is
the left or right shift of a electron in a V-O-V bond.
In another word, the transition is related to the internal charge degrees of freedom in a V-O-V bond.

In this case, a short range order of charge:a dynamical charge disproportionation appears as the instantaneous charge ordering in short range
caused by the coherent left or right shifts of electrons in the neighboring V-O-V bonds.
Therfor, it is very interesting to study the process in which a charge on V-O-V bond disproportionates into different V-sites.
The problem of a dynamical charge disproportionation has been discussed by several authors.\cite{NishimotoE,DamascelliCM}
For example, Damascelli {\it et al.} proposed the charged magnon model to explain the anomaly of optical conductivity
In this model, the occurrence of the asymmetrical charge distribution above {\em T}$_c$ is suggested.
However, the time scale of the dynamical charge disproportionation has not been known yet, since the time window of the most of microscopic probes is much lower
than the time scale of charge hopping.

Recently, it has been proposed by Nishimoto and Ohta that a submillimeter wave ESR is one of the candidates to probe such fast charge motion. \cite{NishimotoE}
As is well known, the ESR line width is a sensitive probe of the spin dynamics.
Although several ESR investigations have been so far reported on NaV$_2$O$_5$.\cite{VasilevE,Schmidt,Yamada}, 
a systematic study of the frequency dependence of ESR line width has not been done yet.
Hence, we examined ESR line width in wide frequency range to study the dynamical charge disproportionation above {\em T}$_c$.
This is the main issue of the present work.

Other interesting subjects in NaV$_2$O$_5$ are the mechanism for the spin gap formation and the related CO pattern.
Although the existence of the CO is widely accepted, the low temperature crystal structure is still unclear.\cite{Ludecke,Sawa}
A ESR measurement gives us useful information to select possible CO patterns as follows.
As reported in the previous paper, we have succeeded in the direct observation of spin gap excitation by means of submillimeter
wave ESR.\cite {Luther}
It was found that the lowest energy gap was located at 8.13 meV and that this mode was triplet, which were consistent with results of neutron
scattering.\cite{Fujii,Yoshihama,Regnault}
To explain the important features such as strong field orientation dependence of ESR intensity, we proposed the existence of the Dzyaloshinsky-Moriya(DM)
interaction in the spin gapped CO phase.
As is well known, the DM interaction appears only in a crystal without an inversion symmetry and the direction of the DM-vector is closely related to the
crystal symmetry.
Therefor, our observation is useful to discuss candidates of CO patterns at low temperature.

\section{Experimental}
  Submillimeter wave ESR measurements have been performed by using pulsed magnetic fields up 30 T.
Far-infrared(FIR) radiations between 95-760 GHz are supplied by an optically pumped FIR laser, Gunn oscillators and backward traveling wave tubes.
We have employed the simple transmission method with Faraday configuration where the propagation vector of the incident radiation is aligned parallel to the
external magnetic field.
Much efforts have been paid to exclude the extrinsic broadening of the line width caused by the inhomogeneity of a magnetic field.
We evaluated the field inhomogeneity by observing the resonance fields of the several small pieces of DPPH located all around the NaV$_2$O$_5$ sample.
Since the line width of each DPPH block is less than 0.1 mT, we can measure the local field at the position of each block by using the sufficiently small size
of DPPH.
By this way, the spatial distribution of the magnetic field over the NaV$_2$O$_5$ sample can be obtained by monitoring the splitting of the resonance
fields among the DPPH blocks set around a NaV$_2$O$_5$ sample.
To minimize the field inhomogeneity, we used a small sample with the dimensions of 0.2$\times$0.5$\times$1.5 mm$^3$ and we adjusted the sample
position within the accuracy of 1 mm in the magnet. 

\section{Results}
\subsection{Frequency dependence of line width}

  The temperature dependence of the ESR spectra at 190 GHz for B${\parallel}$c and the integrated intensity are plotted in Fig. 2 and Fig. 3, respectively.
Above {\em T}$_c$, the signal shows a typical Lorentzian line shape and this fact indicates the existence of a strong exchange narrowing effect. 
Although the resonance field is almost temperature independent as shown in Fig. 2, a characteristic change of the line width is observed.
Above {\em T}$_c$=34 K, the line width decreases as the temperature is decreased.
Such behavior has been observed in many different materials, in which a line width is governed by the anisotropic exchange interaction.\cite{Oshikawa}
The rapid decrease below {\em T}$_c$ is due to the effective dilution of the interactions acting among the spins for the opening of spin gap
and this is associated with an urgent decrease of the intensity as depicted in Fig. 3.
In the figure, the origin of a small residual intensity observed below 20 K may be caused by undimerized V$^{4+}$ spins originated from a small amount of Na
vacancies.
It is also noticed that the width becomes broader below 20 K and the line shape changes from Lorentzian to Gaussian.
As is well known, a Gaussian line shape is expected when exchange narrowing effect and motional narrowing effect are completely lifted.
Thus the changes suggest that the freezing of the dynamical motion of spins associated with the localization of impurity spin-charge.
This point will be discussed later.
In the following, we focus on the signal above {\em T}$_c$.

	The most important result is the remarkable frequency dependence of the line width found below 100 K as shown in Fig. 4.
In the plot, the extrinsic line width caused by the inhomogeneity of a magnetic field is subtracted.
At 95 GHz, the line width decreases linearly as the temperature {\em T} is decreased, which is consistent with the previous report at 9.3 GHz and at 23.5
GHz.\cite{Yamada}. 
At 135 GHz and above 70 K, the line width is almost identical with that at 95 GHz.
However, below 70 K, a definite deviation is observed between data at those two frequencies.
The temperature dependence becomes less steeper as the frequency is increased from 95 GHz to 135 GHz.
As the frequency is increased further, the deviation from the 95 GHz data becomes more pronounced and plateau like structures appear at higher frequencies.
It is noticed that the effect appears below 100 K and the line width is frequency-independent above 100 K.
For this behavior, we speculate that the anomaly of the line width is a precursor effect caused by the dynamical charge disproportionation.

It should be noted that a plateau like structure is also found for the data measured at 134 GHz and 220 GHz by Schmidt {\it {et al.}} \cite{Schmidt}, although the
frequency dependence is masked for the scattering of data. In fact, it was very difficult to suppress the error of the line width at higher frequencies such as
585 GHz or 762 GHz. In Fig. 4, the scattering of the data is less than 0.01 $\%$ of the resonance field but it is finite in the absolute value, especially in very
strong magnetic fields above 20 T.
In spite of such difficulties, the present systematic research elucidates the existence of the remarkable frequency dependence of ESR line width as a
precursor phenomenon of the CO.
It should be noted that the time scale of the dynamics is clearly detected by ESR.

To show the frequency dependence of line width more clearly, the width is plotted as a function of frequency at several different temperatures in Fig. 5.
The data points at 23.5 GHz measured by Yamada {\it et al.} are plotted together.
The plot clearly exhibits, below 100 K, that the frequency dependence of the line width develops gradually towards {\em T}$_c$. 
At {\em T}=36 K, the maximum of the line width is located around 700 GHz, which is related to the typical frequency of this precursor phenomena.
As is well known, the ESR line width is related to the spin dynamics, however, such frequency dependence is not expected for a conventional
quantum spin chain unless the magnetic field becomes comparable with the exchange coupling.
In the present case, even at very high field above 20 T, the field intensity is only a few $\%$ of the saturation field because the
exchange coupling of the present material is very large as {\em J}=560 K. \cite{Isobe} 
Thus the observed frequency dependence cannot be attributed to the field effect of the spin dynamics.
A more plausible origin is the reduction of the exchange narrowing caused by the dynamical charge disproportionation in the V-O-V bonds.
In another word, a slowing down of the charge motion in the V-O-V bonds is detected as the suppression of the exchange narrowing effect.
If this idea is correct,the frequency dependence effect should depend on the polarization of the radiation relative to the V-O-V bonds.

 Figures 6(a) and 6(b) show the temperature dependence of the line width at different frequencies for B${\parallel}$a and for B${\parallel}$b, respectively.
The frequency dependences of the line width are also observed for these configurations.
However, the broadening effect is approximately half of that for B${\parallel}$c.
It is found that the anisotropy between B${\parallel}$a and B${\parallel}$b is small.
This field orientation dependence can be understood as follows.
In the Faraday configuration which we employed for the present work, the polarization of the incident radiation lies in the plane normal to the
external field.
The polarization is random within this plane, because we use the unpolarized light.
It is natural to assume that the coupling between the radiation and the charge motion is strong when a polarization of the incident radiation lies
parallel to the {\em {ab}}-plane, since charges are confined in this plane for the two dimensional network of V-O-V bonds.
Thus the effect is expected to be large for B${\parallel}$c, where the polarization of the incident radiation lies in the {\em {ab}}-plane.
On the other hand, we can expect a smaller effect for B${\parallel}$a and for B${\parallel}$b because only the half of the polarization of the
incident radiation lies in the {\em {ab}}-plane. 
These expected features are consistent with the experimentally found orientation dependence of the line width.
Hence, we consider that our interpretation for the frequency dependence of the line width is plausible.

\subsection{spin-gap excitation}

 In the previous work, we reported the direct observation of the spin gap excitation by ESR.\cite{Luther}
As is well known, the ESR transition between the ground singlet state and the excited triplet state due to the mechanism of magnetic dipole transition is
forbidden in principle.
However, the presence of a non-secular term such as anisotropic exchange interaction or DM-interaction make it possible to observe this transition by means of ESR.
Since these terms are closely related to the symmetry of the crystal, it is very useful to know the symmetry of the low temperature CO phase.
Since the results are given in the previous work, let us show only the most important features of the experiments.
Figure 7 shows the angular dependence of the ESR spectra for three principal crystal axes.
A strong anisotropy of the intensity is found and the absorption is strongest for B${\parallel}$a and is almost zero for B${\parallel}$c.
On the other hand, the intensity of the transition shows no significant field dependence.
These features suggest that the DM-interaction is the origin of the breaking of the ESR selection rule and this point will be discussed in the next
section.\cite{Luther}

The temperature dependence of the spectra is shown in the inset of Fig. 8.
The intensity is also plotted in Fig. 8 as the function of the temperature.
It decreases rapidly as the temperature is increased and the signal disappears above 20 K.
As shown in the inset of Fig. 8, the resonance field of the spectrum shifts to the higher field side with increasing temperature.
The temperature dependence of the energy gap is evaluated from the shifts in the following way.
The ESR signal at this frequency belongs to the {\em S}$_z$=1 branch of the triplet(see Fig. 2 of the reference \cite{Luther}).
In this case, we can expect that the resonance field at a fixed frequency is increased when the zero field energy gap is reduced.
Thus, we can estimate the temperature dependence of the energy gap by using the shift of the resonance field and the result is plotted in Fig. 8.
It was found that the reduction of the gap is only about 1 $\%$ at 20 K.
This small temperature dependence of the energy gap is consistent with the results of neutron scattering.\cite{Fujii}
The present results show that the intensity of the spin gap excitation is much decreased, although the magnitude of energy gap is unchanged.

The rapid decrease of the signal cannot be attributed to the thermal excitation of the spins across the energy gap because the Boltzman factor exp(-{\em
{$\Delta$}}/{\em k}$_B${\em T}) for the lowest triple branch {\em S}$_z$=1 is only 0.009 at 20 T and at 20 K.  
A more plausible interpretation is that the magnitude of the DM-interaction becomes small above 20 K.
As is well known, the DM-interaction appears only in a crystal without inversion symmetry.
As mentioned before, the crystal structure above {\em T}$_c$ is centrosymmetric and thus, the DM-interaction is brought into the system as the consequence of the
CO.
At this point, it is noticed that the intensity of the superlattice peak relating to the CO gradually develops below {\em
T}$_c$ and saturates around 20 K.\cite{Nakao}
In this case, we can expect that the DM-interaction shows a substantial temperature dependence and consequently, we can expect the decrease of the intensity of
the spin gap excitation above 20 K.

Considering the gradual development of the CO below {\em T}$_c$, the broadening of the line width and the change of the line shape from Lorentzian to Gaussian of
the impurity signal, which is mentioned in the former sub-section, can be also explained as follows.
When the CO becomes completely static below 20 K, the impurity charge caused by the Na vacancies cannot hop to different sites and, as a result, the motional
narrowing effect is completely lifted.
It is also noticed that the exchange narrowing effect caused by an exchange coupling between the impurity can be neglected since the density of the impurity is
very small.
As shown above, the present interpretation is more plausible compared to another interpretation that the residual ESR signal is the manifestation of the
antiferromagnetic ordering.\cite{VasilevE}

\section{Discussion}
\subsection{Charge disproportionation and line width}
In this section, we discuss the origin of the frequency dependence of the line width.
Although the general theory of ESR was invented by Kubo and Tomita long time ago\cite{Kubo}, the actual calculation of a correlation
function for a quantum spin chain in the low temperature condition is non-trivial and is still very difficult.\cite{Anderson,Mori,Nagata,Oshikawa}
It is much difficult for NaV$_2$O$_5$ because a charge degree of freedom should be taken into account.
Since no applicable theory has been given for this case, we discuss the results qualitatively in terms of the exchange narrowing effect. 

In the framework of the exchange narrowing theory, the line width $\Delta${\em H} is given by 
\begin{equation}
\Delta H=\frac{\bar{h}}{g\mu _B}\frac{M _2}{\omega _{ex}},
\end{equation}
where {\em M}$_2$ is the second moment of the perturbation and $\omega$$_{ex}$={\em J}$\slash$$\bar{h}$ is the exchange frequency.
The second moment {\em M}$_2$ is caused by the anisotropic part of the spin Hamiltonian such as dipole-dipole interaction, anisotropic exchange(AE) interaction
and DM interaction.
In the following, we do not discuss the origin of {\em M}$_2$.
The main contribution for $\omega$$_{ex}$ is the effective exchange coupling along the {\em {b}}-axis {\em {J}}$_b$.
If the internal charge degrees of freedom in a V-O-V bond is neglected, the system can be mapped onto a {\em S=1/2} spin chain coupled by {\em {J}}$_b$.
We discuss at first this model to show that no frequency-dependence of the line width is expected for this case.

When the frequency of ESR is increased, the resonance field is automatically increased.
Therefor, we have to consider both the frequency effect and the field effect on spin dynamics and on exchange narrowing effect.
As mentioned before, no field effect on spin dynamics is expected for a {\em S=1/2} spin chain as long as the reduced field {\em h}={\em H}\slash{\em
J$_b$} satisfies the condition {\em h}={\em H}\slash{\em J$_b$}$\ll$1.
For the exchange narrowing, since the exchange coupling {\em J$_b$}=560 K is very large in the present system, the condition
$\omega_{ob}\ll\omega_{ex}$ is satisfied in our experiments, where $\omega$$_{ob}$ is the frequency of ESR used in the experiment.
In this case, the frequency effect for the exchange narrowing is negligible.
Another candidate is the field induced staggered field.
This case has been studied intensively for Cubenzoate and a considerable field effect on line width was found.\cite{Oshikawa,Asano}
However, a staggered field is not expected for the crystal structure of NaV$_2$O$_5$.
Moreover, the shift of the resonance field, which should appear for such case, is not observed as shown in Fig. 2 and thus this possibility is also excluded.
Hence, it is turned out that we cannot explain the frequency dependence of line width in the framework of the exchange narrowing effect on a conventional {\em
S=1/2} spin chain. 

When the internal charge degrees of freedom is considered, we can show qualitatively that the dynamical charge disproportionation affects the line width in two
possible mechanisms.
The first candidate is the modulation of the exchange interaction {\em {J}}$_b$ by the charge hopping in a V-O-V bond.
As shown in Fig. 9, for the charge distribution in the bond, several different charge arrangements are possible and the exchange coupling between two
neighboring spins are different among these arrangements.
In this case, the {\em {J}}$_b$ is modulated in the charge hopping frequency $\omega$$_{h}$ and this modulation gives an extra contribution for $\omega$$_{ex}$
in eq(1).
At higher temperature, $\omega$$_{h}$ is much higher than $\omega$$_{ob}$ and thus the modulation of {\em {J}}$_b$ is averaged out.
When temperature is close to {\em T}$_c$ and the dynamic charge disproportionation is well developed, the condition $\omega_{ob}\sim\omega_{h}$ may be satisfied
at high frequencies. 
In this case, the modulation of {\em {J}}$_b$ is decoupled and thus the denominator of the eq (1) is decreased.
As a result, we can expect the increase of the line width at high frequencies when $\omega_{ob}\geq\omega_{h}$.

The second candidate is the instantaneous increase of the second moment {\em M}$_2$.
As is well known, the DM-interaction gives the largest contribution for {\em M}$_2$ and thus we discuss only this case in the following.\cite{Nojiri}
It should be noted that the DM-interaction does not exist in the centrosymmetric crystal structure above {\em T}$_c$ and thus we
have to consider the local symmetry breaking of the crystal to cause the DM-interaction.
When a particular charge arrangement breaks the inversion symmetry of the crystal locally, an anti-symmetrical anisotropy of the exchange
interaction is induced at that position. 
Such anisotropy contributes to the excess increase of the {\em M}$_2$ and thus contributes also to the increase of the line width. 
However, this effect is not detected by ESR when the frequency of the local symmetry breaking is higher than the $\omega_{ob}$ because the effect is
averaged out.
In this way, the increase of the line width is realized only when the condition $\omega_{ob}\geq\omega_{h}$ is satisfied at high frequencies.
At this point, it is important to point out that the symmetrical part of the exchange interaction does not contributes to the {\em M}$_2$.\cite{Oshikawa} 
For this special feature of ESR, the increase of the second moment is realized only when the charge arrangement is associated with the breaking of the local
inversion symmetry.
The possibility of such local symmetry breaking was first proposed by Damascelli {\em et al.} as "charged magnon model" to explain the optical
spectra.\cite{Damascelli}. 
It should be stressed that such local symmetry breaking picture is compatible with our second interpretation.

As discussed above, we can qualitatively explain the frequency dependence of the line width by considering the internal charge degrees of freedom in a
V-O-V bond, although it is not clear which mechanism mainly contributes to the present result.
It is also noticed that the proposed mechanisms are valid for the two-leg ladder model, in which two charges are located in a V-O-V bond.
The essential point is that the dynamical charge distribution can give rise to the modulation or the anistropy of exchange coupling.

A further support for our interpretation is the fact that the precursor effects are observed in many experiments between {\em T}$_c$ and
100 K.\cite{VasilevT,Ravy,Fertey}
This temperature regime is exactly same where we observed the frequency dependence of line width.
Among these experiments, the observation of the diffuse superlattice peak in X-ray scattering\cite{Ravy} attracts much interest, because it is
compatible with our second proposal that the local symmetry breaking of the crystal occurs even above {\em T}$_c$.
Finally, we conclude that the dynamical charge disproportionation is the origin of the frequency dependence of the ESR line width.
It should be stressed that the time scale of the charge hopping is clearly determined, for the first time, by means of high frequency ESR technique.
We also stress here that the our finding is a completely new phenomenon, in a sense that the internal charge hopping in a chemical bond state is examined
experimentally.
 
\subsection{Crystal symmetry of the charged ordered phase}
As is well known, the ESR transition is forbidden between the ground singlet state and the excited triplet state, unless a non-secular term exists in the
spin-Hamiltonian.\cite{Nojiri}
Possible candidates for such non-secular term are the following:(1) DM interaction, (2) staggered field and (3) AE interaction. 
Recently, it has been reported that the crystal structure of the charge ordered phase is acentric and thus the DM interaction should be taken into
account.\cite{Ludecke}

The mechanism (2) and (3) are not adaptable for NaV$_2$O$_5$ for the following reasons. 
For the staggard field mechanism, we can expect two distinct features:(i)the strong field dependence of the intensity and (ii)the absorption is observed for Voight
configuration where {\em k}$\perp${\em B}.
In the present case, these features are not observed.\cite{Luther}
For AE mechanism, the zero field splitting between {\em S}$_z$=1 and {\em S}$_z$=-1 branches is expected and in fact, this feature is experimentally observed
in SrCu$_2$(BO$_3$)$_2$ recently.\cite{NojiriS}
Since no such splitting is observed for NaV$_2$O$_5$(see Fig. 2 of the reference \cite{Luther}), we can also exclude this candidate.
At this point, we mention the theoretical proposal that large AE interaction exist in NaV$_2$O$_5$.\cite{Yaresko}
We point out here that no evidence of large AE is observed in our experiments although the resolution of ESR is very high.

Next we consider the DM interaction. In the previous work, we speculated that the DM-vecotor {\it D} is along the {\it c}-axis from the analogy of the CuGeO$_3$
case. However, for CuGeO$_3$, the situation is rather complicated because there are two inequivalent sites in a unit cell.
Recently, Sakai proposed following selection rules:(i)when ${\em D}$$\parallel${\em B}, ESR is observed for Faraday configuration and the intensity
shows no strong field dependence, (ii)when ${\em D}$$\perp${\em B}, ESR is observed for Voight configuration and the intensity strongly depends on the
field.\cite{Sakai}
Since we measured the ESR in Faraday configuration, we consider the former case.
As shown in Fig. 7, the intensity is strong for ${\em B}$$\parallel${\em a}, weak for ${\em B}$$\parallel${\em b} and zero for ${\em B}$$\parallel${\em c}.
It is also found that the intensity shows no strong field dependence.\cite{Luther} 
Considering these features, we conclude that the {\it D}-vecotor is located in the {\it ab}-plane and is nearly parallel to the {\it a}-axis.

We now discuss the symmetry of the charge ordered phase and the symmetry of the spin-Hamiltonian by using the well known Moriya's criterion about the direction of
the {\em D}-vector.\cite{Moriya}
In the following, we use the notation given by the L${\ddot u}$deck {\it et al.}\cite{Ludecke}
For the two-leg ladder model shown in Fig. 1(b), the DM-interaction exists between neighboring two spins along the {\it a}-axis at V1 sites and possible
direction of the {\it D}-vector are ${\em D}$$\perp${\em c} and ${\em D}$$\perp${\em a}.
For the neighboring two spins along the {\it a}-axis at V2 sites, ${\em D}$$\parallel${\em b} is expected.
For the zig-zag model shown in Fig. 1(c), considering the Moriya's criterion, we can expect
that the DM-interaction exists between two spins in the diagonal direction and ${\em D}$$\perp${\em c}. 
For the in-line model shown in Fig. 1(d), we cannot expect the DM-interaction between the two neighboring spins. 
The former two models are compatible with our proposal that the {\it D}-vecotor is nearly parallel to the {\it a}-axis, however, it is unclear
which model is realized in the present case.
In is also noticed that, in the case of the two-leg ladder model, the dimerization is expected along the {\it a}-axis at V1 sites.
To select the most plausible model, it is necessary to calculate the transition matrix element theoretically, which has not been made yet.
For such calculation should interpret that only the acoustic mode is observed by means of ESR(see Fig. 2
of the reference\cite{Regnault}).

\section{Conclusion}
A remarkable frequency dependence of the ESR line width found above {\em T}$_c$ and is explained as the manifestation of the dynamical charge disproportionation
toward the charge ordering.
The time scale of the charge hopping is clearly determined, for the first time, by means of high frequency ESR technique.
The present results show the important role of the internal charge degrees of freedom in a V-O-V bond.
The direction of the {\em D}-vector is also proposed, which is useful to examine the spin-Hamiltonian and the crystal structure of the charge ordered phase.

\section{Acknowledgements}
We would like to express our gratitude thanks for Prof. Y. Yamada for showing us the data of ESR and for valuable discussions.
We also express our acknowledgements for Prof. N. Ohta, Prof. T. Toyama and Prof. M. Oshikawa for their useful suggestions. 
This work was partially supported by a Grant-in-Aid for Scientific Research from the Ministry of Education, Science, Sports, and Culture of Japan
and by CREST of Japan Science and Technology Corporation.
S. L. acknowledges the support from the Japan Society for the Promotion of Science.

\begin{figure}
\caption{(a)Schematic crystal structure at room temperature projected along the {\em c}-axis.
Arrow and ellipsoide indicate spin and charge distribution, respectively. 
Figures (b)-(d) show possible candidates for charge ordering
patterns:(b)two-ladders, (c)zig-zag and (c)in-line. Large circle and small circle denote V$^{4+}$ and V$^{5+}$ sites,
respectively.} 
\label{autonum}
\end{figure}
\begin{figure}
\caption{Example of ESR spectra for $\em H\parallel\em c$ at 190 GHz. The vertical scale is magnified
by 10 times for upper two traces. The vertical position of each spectrum is shifted for convenience.} 
\label{autonum}
\end{figure}
\begin{figure}
\caption{Temperature dependence of the integrated intensity at 190 GHz for $\em H\parallel\em c$.} 
\label{autonum}
\end{figure}
\begin{figure}
\caption{Temperature dependence of line width for $\em H\parallel\em c$ at different frequencies} 
\label{autonum}
\end{figure}
\begin{figure}
\caption{Frequency dependence of line width for $\em H\parallel\em c$. The data at 23.5 GHz are identical with those plotted in Fig. 1 of
Ref.[22]. Lines are eye-guides.} 
\label{autonum}
\end{figure}
\begin{figure}
\caption{Temperature dependence of line width for (a)$\em H\parallel\em a$ and (b)$\em H\parallel\em
b$ at different frequencies} 
\label{autonum}
\end{figure}
\begin{figure}
\caption{ESR spectra for spin gap excitation measured at 2.52 THz for three different field orientations. Vertical scale is magnified for
lower two traces.} 
\label{autonum}
\end{figure}
\begin{figure}
\caption{Temperature dependence of the intensity of spin gap excitation signal and temperature dependence of the
energy gap evaluated from the shift of the resonance field by the method mentioned in the text. The inset shows the temperature dependence of ESR spectra at 2.52
THz for$\em H\parallel\em a$} 
\label{autonum}
\end{figure}
\begin{figure}
\caption{Schematic views for several possible charge distribution. Ellipsoid and gourd-shaped area show symmetrical and asymmetrical charge distributions in a
V-O-V bond, respectively. Exchange coupling along the leg of the ladder are different among different charge distribution patterns and a local breaking of the
inversion symmetry is caused for a particular type of charge distribution.} 
\label{autonum}
\end{figure}


\begin{references}
\bibitem{Isobe}M. Isobe and Y. Ueda, J. Phys. Soc. Jpn. {\bf 65,} 1178(1996).
\bibitem {Carpy}P. A. Carpy and J. Galy, Acta Cryst. {\bf B 31,} 1481(1975).
\bibitem{Smolinski}H. Smolinski, C. Gros, W. Weber, U. Peuchert, G. roth, M. Weiden and C. Geibel, Phys. Rev. Lett.
{\bf 80,} 5164(1998).
\bibitem{Meetsma}A. Meetsma, J. L. de Boer, A. Damascelli, J. Jegoudez, A. Revcolevschi and T. T. M. Palstra, Acta. Cryst.
 {\bf C54,} 1558(1998).
\bibitem {Schnering}H. G. von Schnering, Yu. Grin, M. Kaupp, M. Somer, R. K. Kremer and O. Jepsen, Z. Krystallogr
 {\bf 213,} 246(1998). 
\bibitem{Seo}H. Seo and H. Fukuyama, J. Phys. Soc. Jpn. {\bf 67,} 2602 (1998).
\bibitem{Thalmeier}P. Thalmeier and P. Fulde, Europhys. Lett. {\bf 44,} 242(1998).
\bibitem {Nishimoto}S. Nishimoto and Y. Ohta, J. Phys. Soc. Jpn. {\bf 67,}  3679(1998).
\bibitem{Mostovoy}M. Mostovoy and D. Khomskii, cond-mat/9806125.
\bibitem{Gros}C. Gros and R. Valenti, Phys. Rev. Lett. {\bf 82,} 976(1999).
\bibitem{Ohama}T. Ohama, H. Yasuoka, M. Isobe and Y. Ueda, Phys. Rev. B {\bf 59,} 3299(1999).
\bibitem {VasilevT}A. N. Vasil'ev, V. V. Pryadun, D. I. Khomskii, G. Dhalenne and A. Revcolevschi, M. Isobe and Y. Ueda, Phys. Rev. B {\bf 56,} 5065(1997).
\bibitem{Smirnov}A. I. Smirnov, M. N. Popova, A. B. Sushkov, S. A. Golubchik, D. I. Khomskii, M. Mostovoy, A. N. Vasil'ev, M. Isobe and
Y. Ueda, Phys. Rev. B {\bf 59,} 14546 (1999)
\bibitem{Sekine}Y. Sekine, N. Takeshita, N. Mori, M. Isobe and Y. Ueda, unpublished
\bibitem{Ravy}S. Ravy, J. Jegoudez and A. Revcolevschi, Phys. Rev. B {\bf 59,} R681(1999).
\bibitem{Damascelli}A. Damascelli, D. van der Marel, M. Gruninger, C. Presura, T. T. M. Palstra, J. Jegoudez and A.
Revcolevschi, Phys. Rev. Lett. {\bf 81,} 918(1998). 
\bibitem{Fertey}B. Fertey, M. Poirier, M. Castonguay, J. Jegoudez and A. Revcolevschi, Phys. Rev. B  {\bf 57, } 13698(1998).
\bibitem{NishimotoE}S. Nishimoto and Y. Ohta, J. Phys. Soc. Jpn. {\bf 67,}  4010(1998).
\bibitem{DamascelliCM}A. Damascelli,  C. Presura, D. van der Marel, J. Jegoudez and A.
Revcolevschi, con-dmat/9906042.
\bibitem {VasilevE}A. N. Vasil'ev, A. I. Smirnov, M. Isobe and Y. Ueda, Phys. Rev. B {\bf 56,}  5065(1997).
\bibitem {Schmidt}S. Schmidt, W. Palme, B. Luthi, M. Weiden, R. Hauptmann and C. Geibel, Phys. Rev. B {\bf 57,}  2687(1998).
\bibitem {Yamada}I. Yamada, H. Manaka, H. Sawa, M. Nishi, M. Isobe and Y. Ueda, J. Phys. Soc. Jpn. {\bf 67,}  4269(1998).
\bibitem{Ludecke}J. L${\ddot u}$decke, A. Jobst, S. van Smaalen, E. Morr${\acute e}$, C. Geibel and H. G. Krane, Phys. Rev. Lett. {\bf 82,}  3633(1999).
\bibitem{Sawa}Sawa {\it {et al.}}, unpublished
\bibitem {Luther}S. Luther, H. Nojiri, M. Motokawa, M. Isobe and Y. Ueda, J. Phys. Soc. Jpn. {\bf 67,} 3715(1998).
\bibitem{Fujii}Y. Fujii, H. Nakano, T. Yoshihama, M. Nishi, K. Nakajima, K. Kakurai, M. Isobe, Y. Ueda and H. Sawa, J. Phys. Soc. Jpn.
{\bf 66,} 326(1997).
\bibitem{Yoshihama}T. Yoshihama, M. Nishi, K. Nakajima, K. Kakurai, Y. Fujii, M. Isobe, C. Kagami and Y. Ueda, J. Phys. Soc. Jpn.
{\bf 67, } 744(1998).
\bibitem{Regnault}L. P. Regnault, J. E. Lorenzo, J. P. Boucher, B. Grenier, A. Hiess, T. Chatterji J. Jegoudez and A.
Revcolevschi, to be published in Physica B.
\bibitem{Nakao}H. Nakao, K. Ohwada, N. Takesue, Y. Fujii, M. Isobe, Y. Ueda, H. Sawa, H. Kawada, Y. Murakami, W. I. F. David, R. M. Ibberson, Physica B  {\bf
241-243,} 534 (1998). 
\bibitem {Kubo}R. Kubo and K. Tomita, J. Phys. Soc. Jpn. {\bf 9,} 1954(91).
\bibitem {Anderson}P. W. Anderson, J. Phys. Soc. Jpn. {\bf 9,} 316(1954).
\bibitem {Mori}H. Mori and K. Kawasaki, Prog. Theor. Phys.  {\bf 27,} 529(1962).
{\it {ibid.}} {\bf 28,} 971(1962).
\bibitem {Nagata}K. Nagata and Y. Tazuke, J. Phys. Soc. Jpn.  {\bf 32,} 337(1972).
\bibitem {Oshikawa}M. Oshikawa and I. Affleck, Phys. Rev. Lett. {\bf 82,} 5136(1999).
\bibitem{Asano}T. Asano, H. Nojiri, Y. Inagaki, J. P. Boucher, T. Sakon, Y. Ajiro and M. Motokawa, cond-mat/9911094
\bibitem {Nojiri}H. Nojiri, H. Nojiri, H. Ohta, S. Okubo, O. Fujita, J. Akimitsu, and M. Motokawa, J. Phys. Soc. Jpn. {\bf 68} (1999) 3417.
\bibitem {NojiriS} H. Nojiri, H. Kageyama, K. Onizuka, Y. Ueda and M. Motokawa, J. Phys. Soc. Jpn. {\bf 68,} (1999)2906.
\bibitem{Yaresko}P. Thalmeier and A. N. Yaresko, cond-mat/9904443
\bibitem{Sakai}T. Sakai, private communication
\bibitem {Moriya}T. Moriya, Phys. Rev. {\bf 120,} 91(1960).




\end{references}
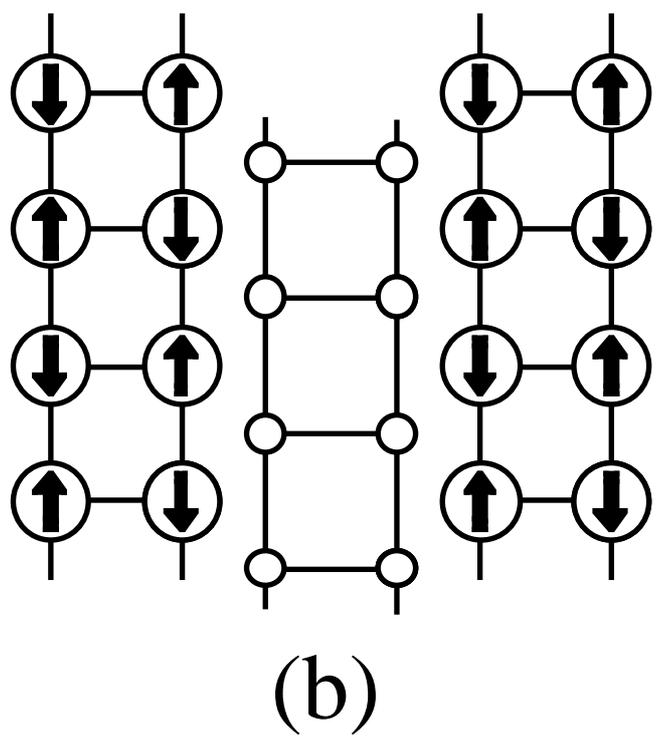
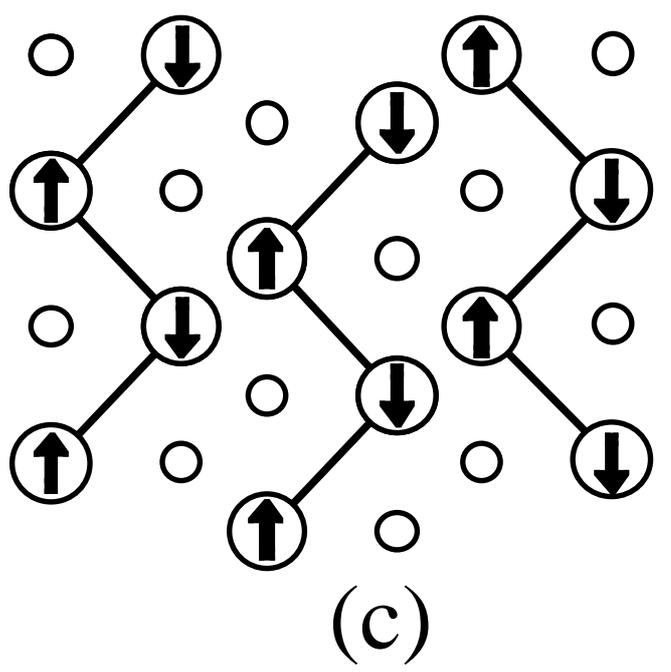
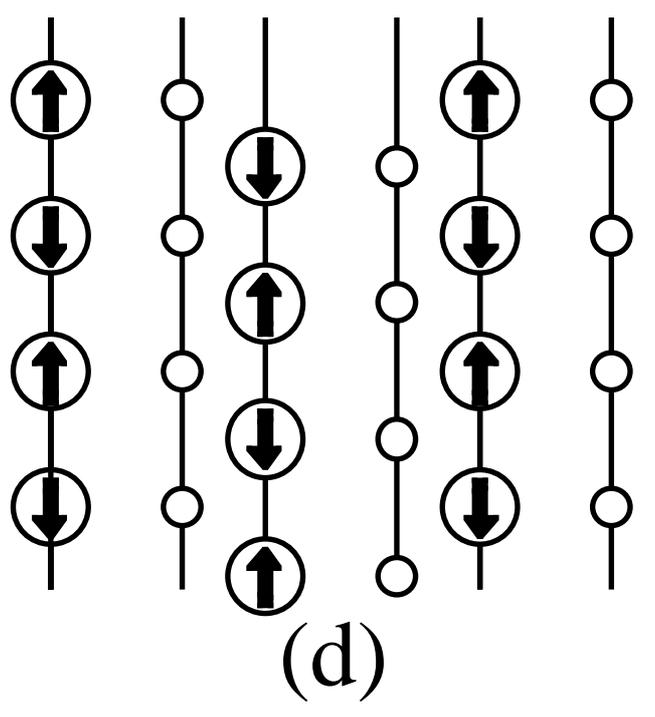
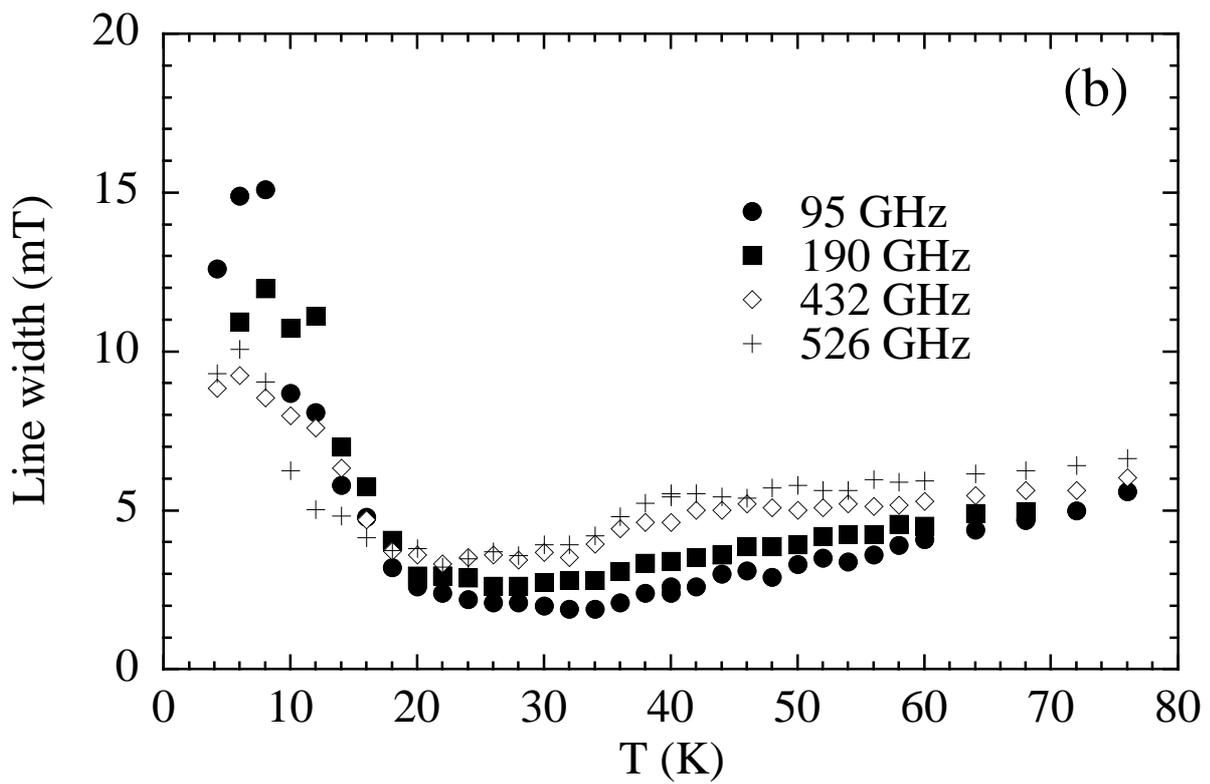
\end{document}